\documentclass[manuscript,nonacm]{acmart}

\AtBeginDocument{%
  }

\setcopyright{acmlicensed}
\copyrightyear{2025}
\acmYear{2025}
\acmDOI{XXXXXXX.XXXXXXX}

\acmJournal{JACM}
\acmVolume{37}
\acmNumber{4}
\acmArticle{111}
\acmMonth{8}

\begin{document}

\title{The Agentic Economy}

\author{David M. Rothschild}
\authornote{Corresponding author is David M. Rothschild, \texttt{davidmr@microsoft.com}.}
\author{Markus Mobius}
\author{Jake M. Hofman}
\author{Eleanor Dillon}
\author{Daniel G. Goldstein}
\author{Nicole Immorlica}
\author{Sonia Jaffe}
\author{Brendan Lucier}
\author{Aleksandrs Slivkins}
\author{Matthew Vogel}

\affiliation{
  \institution{\\ Microsoft Research}
  \country{USA}
}

\authorsaddresses{} %

\renewcommand{\shortauthors}{Rothschild, Mobius, Hofman, Dillon, Goldstein, Immorlica, Jaffe, Lucier, Slivkins, and Vogel} %



\begin{abstract}
Generative AI has transformed human-computer interaction by enabling natural language interfaces and the emergence of autonomous agents capable of acting on users' behalf. While early applications have improved individual productivity, these gains have largely been confined to predefined tasks within existing workflows. We argue that the more profound economic impact lies in reducing communication frictions between consumers and businesses. This shift could reorganize markets, redistribute power, and catalyze the creation of new products and services. We explore the implications of an agentic economy, where assistant and service agents interact programmatically to facilitate transactions. A key distinction we draw is between unscripted interactions—enabled by technical advances in natural language and protocol design—and unrestricted interactions, which depend on market structures and governance. We examine the current limitations of siloed and end-to-end agents, and explore future scenarios shaped by technical standards and market dynamics. These include the potential tension between agentic walled gardens and an open web of agents, implications for advertising and discovery, the evolution of micro-transactions, and the unbundling and rebundling of digital goods. Ultimately, we argue that the architecture of agentic communication will determine the extent to which generative AI democratizes access to economic opportunity.
\end{abstract}

\maketitle

\section{Introduction}
Generative AI has revolutionized the way we interact with technology, allowing people to express their intent in free-form natural language. It has paved the way for AI agents that not only converse with users but also perform actions on their behalf, flexibly and with minimal guidance. 
Delegation to AI has already begun to improve the efficiency of \textit{individual processes}, making both consumers and businesses more productive in the set of tasks they had already been doing~\cite[e.g.,][]{brynjolfsson2025generative,cui2024effects}. However, we believe that the more disruptive---and yet to be realized---impact of generative AI is its potential to drastically reduce the communication frictions \textit{between (and among) consumers and businesses}. This could lead to a reorganization of markets, shifts of market power, and the introduction of entirely new products and services.

Consumers have traditionally faced high communication costs when initiating relationships with businesses, reducing efficiency~\cite{klemperer1995competition}. For example, a consumer seeking a new tax preparer might hesitate to switch because she would have to explain her financial situation all over again to a new person or online service. These communication hurdles can prevent consumers from taking advantage of better products and services or lower prices. Businesses have tried to lower these costs with tools like online forms and voicemail menus, but these often just shift communication costs to the consumer and can make interactions more rigid.

Imagine instead a future where every consumer has an \textit{assistant agent} to communicate their preferences and personal information to businesses, and every business has \textit{service agents} to interact with consumers and other businesses. These agents could be designed to interface with each other seamlessly and flexibly, transforming the landscape of consumer-business interactions. Delegating interactions to such assistant and service agents lowers communication costs, expanding the landscape of options available to both consumers and businesses, increasing the efficiency of markets.

To unlock the full economic potential of generative AI’s communication capabilities, two developments are necessary. First, consumers and businesses must widely adopt assistant and service agents. This is already underway~\cite{bick2024rapid,humlum2024adoption}. Second, these agents must be designed to programmatically interact with each other to facilitate transactions. On the technical front, there has been significant progress in standardizing such agentic interaction, with frameworks like Microsoft's AutoGen \cite{MicrosoftAutogen}, and protocols like Anthropic's Model Context Protocol \cite{AnthropicMCP} and Google's Agent2Agent Protocol \cite{GoogleA2A}. However, it remains to be seen how these advances will be adopted and implemented, or constrained, given their complex interplay with and dependence on market forces.

The largest benefits of inter-agent communication will be realized as markets reorganize around these new capabilities  \cite{BRESNAHAN1995}.  While the possibilities of this technological distribution are numerous, we offer a framework for understanding the most plausible outcomes and contrasting them with the current state. The key question is whether inter-agent communication will occur within closed "agentic walled gardens" controlled by a few dominant providers, akin to today's app stores, or through a more open "web of agents" where agents freely connect and transact, much like today's World Wide Web. The answer to this question will determine how today's largest online platforms are impacted by the proliferation of interconnected agents, whether it be further entrenchment or diminished market dominance and democratization of the economic benefits of AI technology. We explore 
how those benefits might manifest, with implications for advertising, e-commerce, and the creation of new products and industries.

\section{The Current State of Agentic AI}

Before discussing future possibilities for agentic economies, it is helpful to survey the current landscape of AI agents.
On the surface it may appear as if several existing efforts are well on their way to providing consumers and businesses with assistant and service agents that could function as described above.
But most existing agents lack a key ingredient: while they are designed to interact with or simulate human users, few public offerings are designed to interact with each other.\looseness=-1

Existing agents generally come in one of two forms: siloed service agents or general-purpose end-to-end agents.
The first, siloed service agents, provide a new user interface for products and services within a single company.
For example, Amazon's Rufus allows customers visiting Amazon to access their order histories or compare potential purchases by features through natural language instead of navigating a website. 
Likewise, Expedia’s Romie provides a chat interface to help customers build travel itineraries---including flights, hotels, and restaurants---by pulling information from customer emails and group chats.
Importantly, however, these efforts do not expose interfaces intended for interaction with 
other agents.
As a result, they are still fundamentally limited by people having to navigate to and personally interact with them.

End-to-end agents take a different approach, aiming to provide general-purpose automated functionality that is not limited to a single company or service.
For instance, technology from OpenAI, Google, and Microsoft 
can aggregate research from external sources, navigate business websites on the user's behalf, and even perform simple tasks like making reservations or placing food orders, all within one interface.
Importantly, however, much of the functionality that these end-to-end agents provide currently comes through ``computer use models'' that simulate a user pointing and clicking on existing (non-agentic) websites.
This gives the illusion of assistant and service agents working together, but the absence of a true service agent on the business side limits what an end-to-end agent can accomplish. Moreover, by mimicking human users, these agents risk creating adversarial relationships with businesses that, for example, rely heavily on advertising revenue and may resist having their websites accessed by agents instead of humans.
Even if end-to-end agents can perfectly simulate human users and businesses agree to agent-based access, interactions would still be constrained by what businesses currently expose through existing web forms
that limit the expressiveness of requests and responses. 
For instance, businesses offering highly customizable products---such as catering services---often use only basic contact forms, handling niche or specialized requests through human follow-up.

\section{The Future Impact of Agentic AI}
Given their limitations, we anticipate that siloed and end-to-end agents will ultimately give way to 
agents that are designed to seamlessly interact directly with each other.
However, there is a complex interplay between technical capabilities and market forces that could lead to a number of different scenarios for what this solution will look like, who will control it, and what it will be capable of.

\subsection{The Market Power of Digital Intermediaries}
Two-sided platforms (a businesses designed to bring together two distinct sides for a transaction) such as Amazon, Expedia, OpenTable, and Spotify are key intermediaries of the current digital economy that create value by matching millions of consumers and businesses to each other within specific domains such as shopping, travel, dining, and music~\cite{rysman2009economics}. 
They do this in part by standardizing how both sides interact. For example, Amazon requires sellers to follow specific formats and policies, while consumers must use its interface to search and transact. In exchange, consumers access a vast range of sellers, and businesses gain exposure to a large customer base. But this comes with trade-offs: both operate within a tightly constrained system, are subject to the platform's design choices (e.g., ranking algorithms), and pay referral fees.

If an agentic economy enables each consumer's assistant agent to directly and flexibly communicate with each businesses' service agent via \textit{unscripted communication}, the role---and market power---of intermediary platforms could shift substantially.
In principle, once communication frictions are low enough, interoperable AI agents could eliminate the necessity for two-sided platforms as intermediaries all together.
Consumer assistant agents could directly find and flexibly negotiate with service agents (\cite{davidson2024evaluating}) to buy goods, book hotels and airlines, make dining reservations, and stream music.
This would represent a drastic decentralization of power compared to today's markets.

In practice, however, intermediary platforms often provide value beyond simply standardizing communication between buyers and sellers, via discovery, validation, remediation, and economies of scale. For instance, in domains like travel, it may still be useful for an intermediary to provide trusted suggestions, trip insurance, dispute resolutions, or ensure regulatory compliance. However, an agentic economy could lead to fierce competition between intermediaries due to low switching costs, reducing the profit they can extract.

\subsection{Agentic Walled Gardens vs. the Web of Agents}

Even though any specific assistant and service agent will be technically capable of communicating with each other in an unscripted manner, they might be \textit{restricted} with whom they are allowed to interact with due to market forces. Select firms may provide assistant agents for free but restrict communications, creating ``agentic walled gardens.’’ In some sense this would be a natural evolution of today's existing application ecosystems such as the app stores in dominant operating systems.
Given their existing large user bases and nascent assistant technologies like Apple Intelligence, Google Assistant, Microsoft Copilot, and Meta AI, these firms are well-positioned to extend their current marketplaces to include interoperable AI agents.
For example, in March 2025, Meta launched basic service agents for business pages on Facebook and Instagram at no cost, but these service agents are only accessible to users on their own platforms. In addition, firms such as OpenAI and Anthropic that have built out large user bases for their assistant agents could develop their own marketplaces.

These walled gardens could offer a number of benefits such as ensuring a baseline of quality and security by filtering out low-quality or fraudulent service agents and streamlining discoverability and rating of agents; they could also offer insurance if an agent makes a mistake. However, much like today’s app stores, this could concentrate market power in the hands of a few dominant players, who could leverage their position to extract substantial profits and limit the openness, competitiveness, and innovation of the broader ecosystem.
This could also lead to fragmentation of agentic ecosystems and suboptimal user experiences---for instance, a given individual might have siloed assistant agents for personal and professional purposes, making it difficult to coordinate between them.

Conversely, if consumers and businesses fully own and manage their agents, communication could be both unscripted and unrestricted, leading to a completely open and decentralized ``web of agents'' that is not controlled by any one entity. Similar to today's World Wide Web, any agent could join and transact with any other. Assistant agents would play a role similar to web browsers, and service agents similar to websites. This could foster competition, innovation, and broad access to agentic technology, but it also comes with substantial challenges.
In particular, successful development of a web of agents requires large-scale coordination among many players---including corporations and governments---to develop and agree upon standards and protocols. It also requires robust mechanisms for discovery, trust, and security among interacting agents. 

\subsection{The Future of Advertising}
In today's digital economy there are generally many more businesses and products available to consumers than they have the time to research and consider, and so advertising plays a crucial role in capturing attention and guiding online transactions.
But in an agentic economy where assistants can interact with millions of businesses on behalf of consumers, attention is a less constrained resource. What matters more is the algorithm that matches assistants to service agents. 

In a scenario where there are strong central intermediaries (i.e., an agentic walled garden or web of agents with dominant discovery layers) some form of paid prioritization, akin to today’s advertising, is very likely to influence rankings. But the truly scarce and valuable resource---particularly, in a web of agents---will be high-quality human feedback on goods and services. This feedback will be crucial not only for improving offerings (e.g., training AI agents) but also for distinguishing high-quality services from poor ones.  The focus of monetization and competition could shift from the “attention economy” to a “preference economy.” Success will hinge on attracting early, engaged users who provide valuable feedback.

New means of monetization may emerge to aggregate and rank service agents based on feedback from consumers and their assistants. 
Just as people leave reviews on platforms like Yelp or Google Maps today, future assistants might generate reviews based on user satisfaction data. Businesses will compete by offering better prices or services to attract these early users, creating a flywheel of feedback and preference data.

\subsection{Payments and Micro-transactions}

As interactions between consumers and businesses become more seamless and platform intermediaries less central, we anticipate a rise in “one-off” transactions and accompanying decrease in repeat engagements and long-term consumer-business relationships. This shift may encourage the growth of micro-transactions. For example, a consumer whose assistant frequently and seamlessly switches between multiple content services (two-sided platforms) ---Spotify and Pandora, or Netflix and Amazon Prime---may prefer usage-based micro-payments over subscribing to both services.  Furthermore, the usual hassle cost of micro-payments is rendered moot when transactions are handled entirely by assistants and service agents.  This  further encourages the use of micro-transactions that would otherwise be too inconvenient to handle manually. This is likely to evolve regardless of whether walled gardens of web of agents scenario triumphs.

\subsection{Unbundling, Rebundling, and New Products}

Products, goods, and services are often bundled to balance complexity and efficiency of transactions. For example, a news article may bundle interviews, photos, and facts, even if the reader is already familiar with parts of the story. A consumer's assistant agent, however, can track what the user has already read and collaborate with a service agent (e.g., from the New York Times) to generate a customized article focusing only on new or relevant information. This dynamic and personalized rebundling optimizes knowledge transfer while minimizing cognitive load. The interaction between assistant and service agents could spark a new market for digital data sources to be used in the generation of personalized content by AI agents, such as through Retrieval-Augmented Generation (RAG). In principle, an assistant agent could pull from multiple sources of high-quality content to create customized offerings for its user.  Currently, however, RAG is often used with public-domain sources like Wikipedia or under fair use claims, as publishers are generally unwilling to permit their content to be reused without compensation. This limits the full potential of generative AI to produce personalized content for users. Through micro-transactions for the use of individual pieces of digital content, assistants and service agents could enable a RAG ecosystem that compensates content creators while unlocking the ability to create customized user experiences.

More generally, the power of dynamic and personalized bundling could apply to many other digital goods and services that can be deconstructed and reconstructed by assistant agents working with service agents to best serve the needs of end users.  This potential for feature bundling synergizes directly with our anticipated rise of micro-transactions handled by agents.  As the infrastructure for micro-payments develops, it will become possible for digital components to be individually negotiated, priced, and packaged into hyper-personalized products on the fly. We expect more extreme unbundling, rebundling, and new products in a webs of agents where there can be unrestricted communication between the assistant any \textit{any} service agent, allowing for more innovation in product creation and bundles.\looseness=-1

\section{Conclusion}

As technological progress drives greater specialization, it also increases the need for coordination, which in turn demands more sophisticated communication between individuals, organizations, and systems \cite{10.2307/2118383}. For example, where one doctor once handled a patient’s care, today multiple specialists must collaborate using advanced tools and shared information. AI agents mark a shift in this pattern: rather than compounding communication overhead, they offer a way to streamline it. By coordinating tasks across fragmented systems, agents stand to reduce friction in markets, lower switching costs, and unlock more decentralized access to digital goods and services.

Yet the architecture of this emerging marketplace is still taking shape. Much will depend on whether AI agents are allowed to interact freely across an open web or whether their interactions are restricted within closed ecosystems. These paths will be shaped by emerging technical standards, evolving regulations, and of course the choices made by early stakeholders.
In that sense, this moment is reminiscent of the early days of the Internet before it began generating billions of dollars of revenue. The current platform-dominated economy was certainly not obvious in the late 1990s into the early 2000s, when  digital commerce was becoming increasingly decentralized as early Internet portals gave way to a blossoming World Wide Web.

Now is the time for us to reflect and ask what kind of agentic economy we want in the near future. The choices we make today will determine not only how these markets function, but also who benefits from this new wave of technology.

\bibliographystyle{unsrt}
\bibliography{bibliography}

\end{document}